\documentclass[aps,pre,superscriptaddress,twocolumn]{revtex4-1}

\usepackage{amsmath}
\usepackage{float}
\usepackage{amsfonts}
\usepackage{graphicx}
\usepackage{xcolor}

\begin{document}


\title{Small-scale properties of a stochastic cubic-autocatalytic reaction-diffusion model}


\author{Jean-S\'{e}bastien Gagnon}
\email[]{gagnon01@fas.harvard.edu}
\affiliation{Department of Earth and Planetary Sciences, Harvard University, Cambridge, Massachusetts, USA}

\author{David Hochberg}
\email[]{hochbergd@cab.inta-csic.es}
\affiliation{Centro de Astrobiolog\'{i}a (CSIC-INTA), Madrid, Spain}

\author{Juan P\'{e}rez-Mercader}
\email[]{jperezmercader@fas.harvard.edu}
\affiliation{Department of Earth and Planetary Sciences, Harvard University, Cambridge, Massachusetts, USA}
\affiliation{Santa Fe Institute, Santa Fe, New Mexico, USA}


\date{\today}

\begin{abstract}
We investigate the small-scale properties of a stochastic cubic-autocatalytic reaction-diffusion (CARD) model using renormalization techniques.  We renormalize noise-induced ultraviolet divergences and obtain beta functions for the decay rate and coupling at one-loop.  Assuming colored (power law) noise, our results show that the behavior of both decay rate and coupling with scale depends crucially on the noise exponent.  Interpreting the CARD model as a proxy for a (very simple) living system, our results suggest that power law correlations in environmental fluctuations can both decrease or increase the growth of structures at smaller scales.
\end{abstract}

\pacs{}

\maketitle


\section{Introduction \label{sec:Introduction}}

The macroscopic (or large scale) behavior displayed by complex systems is, in general, the result of complicated mechanisms operating at shorter space-time scales.  For instance, an eukaryotic cell contains a nucleus and a myriad of organelles that interact via numerous chemical nonlinear interactions to produce complex macroscopic behaviors (like movement and replication).  Together, these mechanisms make ``the whole greater than the sum of its parts''.

In many instances, we only have access to a large scale dynamical description of the system and, of course, we are interested in determining what is the underlying internal {\em mechanism} that makes the system work at the larger scales. Answering the above question is in general not possible and in the cases where the question can be formulated, answering it is a daunting task. However, given the ubiquity and relevance of these systems, it is worthwhile to try to gain insight from simpler, more tractable models.

In this context we can ask if, for example, there are generic features of these systems that we could use to our advantage when trying to go from large (or infrared, IR) to short (or ultraviolet, UV) scales. Fortunately, from the study of chemistry and out-of-equilibrium systems, it is known that the IR-scale dynamics of many complex systems contains, at least, three basic components: a diffusive component, a reactive component and a stochastic component.  Broadly, they represent the exchange of chemical information, its processing and the influence of the environment (both internal and external).

The goal of this paper is to study a prototype of chemical system possessing the above three characteristics.  We focus on the stochastic version~\cite{Lesmes_etal_2003} of a two-species chemical model originally introduced by Higgins~\cite{Higgins_1964} and Selkov~\cite{Selkov_1968} in their studies of glycolysis, and later on used by Gray and Scott~\cite{Gray_Scott_1983,Gray_Scott_1984,Gray_Scott_1985} (see also Ref.~\cite{Prigogine_Nicolis_1977}) as a model for autocatalysis.  This cubic autocatalytic reaction-diffusion (CARD) model has an interesting phenomenology.  Indeed, numerical simulations of the deterministic CARD model show the appearance of a variety of spatiotemporal patterns such as stripes, spirals and self-replicating domains~\cite{Pearson_1993} (see Ref.~\cite{Mazin_etal_1996} for a bifurcation analysis).  Self-replicating domains have been observed experimentally in a chemical system and can be qualitatively explained using the CARD model~\cite{Lee_etal_1993,Lee_etal_1994}.  This makes the CARD model a useful proxy for a (very simple) living system (see Sect.~\ref{sec:Model_properties}).  Understanding the behavior of such a simple system might provide insight into real, more complex organisms.

In a more realistic setting, the effect of the environment is introduced via a noise term.  The same types of patterns are obtained in simulations of the stochastic CARD model~\cite{Lesmes_etal_2003,Zorzano_etal_2004} compared to its deterministic counterpart, with the important difference that the noise strength can be varied to induce transitions between patterns.  This kind of noise-controlled transition has been observed experimentally in a chemical system, where switching between two types of mechanical noise (stirring and shaking) leads the system to evolve following two different chemical pathways~\cite{Carnall_etal_2010}.  Similarly, clockwise and counterclockwise stirring has been shown to induce chirality in certain chemical solutions \cite{Ribo_etal_2001} (see also Ref.~\cite{Hochberg_2009} for the use of noise to model the effect of planetary conditions on chemical homochirality production).  Noise-induced oscillations have also been observed in the Belousov-Zhabotinski reaction~\cite{Simakov_PerezMercader_2013}.  This shows that noise can play an important part in the evolution of a chemical system.  Conversely, it is conceivable to use externally tunable noise to probe the microscopic dynamics of such a system.

A previous theoretical study of the stochastic CARD model can be found in Ref.~\cite{Hochberg_etal_2003}.  There, the authors use Wilsonian renormalization group techniques to coarse-grain the model and study its dynamics at large scales.  In this paper, we are interested in going from larger to smaller scales in order to shed some light on internal structures and chemical reactions taking place inside domains.  In other words, we want to fine-grain the system, as opposed to the well-known coarse-graining in condensed matter physics.  To do that we use the renormalization group, but run it towards the UV.

As a first step, we investigate the appearance of nontrivial behavior at small scales in this simple model.  For that, we compute the effects of fluctuations by renormalizing noise-induced divergences in order to obtain beta functions.  This step, and the intricacies encountered along the way (such as regularization in the presence of noise), will set the stage for future work along those lines.
 
Note that the stochastic CARD model is a particular example of a system of stochastic partial differential (SPDE) equations.  Although we focus on this model, the tools developed here could in principle be applied to other systems.  Phenomenologically interesting examples include reaction-diffusion chemical models (such as the Oregonator~\cite{Field_Noyes_1974} and the Brusselator~\cite{Prigogine_Nicolis_1977}), the Kardar-Parisi-Zhang (KPZ) equation~\cite{Kardar_etal_1986} and the Burgers equation~\cite{Forster_etal_1977,Medina_etal_1989}.

The rest of this paper is organized as follows.  We present the stochastic CARD model and its properties (with particular emphasis on the noise) in Sec.~\ref{sec:Gray_Scott_model}.  Section~\ref{sec:Renormalization} deals with the details of the beta function computations for the parameters of the CARD model.  Our results for the renormalization analysis (fixed points and flow diagrams) are presented and discussed in Sect.~\ref{sec:Discussion}.  We conclude in Sect.~\ref{sec:Conclusion}.  Feynman rules and other technical details of one-loop computations (including regularization in the presence of noise) are relegated to appendices.

\section{The stochastic CARD model \label{sec:Gray_Scott_model}}

\subsection{Model and properties}
\label{sec:Model_properties}

In the following we consider a stochastic version of the CARD model~\cite{Lesmes_etal_2003}.  The CARD model is based on the following model chemical reactions:
\begin{eqnarray}
\label{eq:Gray_Scott_reactions}
\mathrm{U} + 2\mathrm{V} & \stackrel{\lambda}{\rightarrow} & 3\mathrm{V}, \nonumber \\
\mathrm{V} & \stackrel{r_{v}}{\rightarrow} & \mathrm{P}, \nonumber \\
\mathrm{U} & \stackrel{r_{u}}{\rightarrow} & \mathrm{Q}, \nonumber \\
  & \stackrel{f}{\rightarrow} & \mathrm{U}.
\end{eqnarray}
The substrate U is fed into the system at a constant rate.  The species V turns the substrate U into V via an autocatalytic reaction.  Both species U and V undergo an irreversible reaction and become the inert products P and Q.  In the interpretation that the stochastic CARD model can be taken as a proxy for a living system, the substrate U can be viewed as food continuously fed into the system.  The species V is the organism itself, forming domains (or ``cells'') over the substrate via some form of metabolism embodied by the autocatalytic reaction.  

The evolution equations corresponding to the Gray-Scott reactions~(\ref{eq:Gray_Scott_reactions}) in the general case where diffusion and noise are present are:
\begin{eqnarray}
\label{eq:Gray_Scott_equations_1}
\frac{\partial V}{\partial t} & = & D_{v}\nabla^{2}V - r_{v}V + \lambda UV^{2} + \eta_{v}(x), \\
\label{eq:Gray_Scott_equations_2}
\frac{\partial U}{\partial t} & = & D_{u}\nabla^{2}U  - r_{u}U - \lambda UV^{2} + \eta_{u}(x) + f,
\end{eqnarray}
where we use the shortcut notation $x = ({\bf x},t)$, $B=B({\bf x},t)$ is the spacetime dependent concentration for species B (with B = U, V), $D_{b}$ is the diffusion constant for species B, $r_{b}$ is the decay rate into inert products for species B, $\eta_{b}(x)$ is the spacetime dependent stochastic noise term for species B, $\lambda$ is the rate constant for the autocatalytic reaction between U and V and $f$ is the constant feed rate of U into the system.  All model parameters are positive.  Note that we use roman script to denote chemical compounds and italic script for concentrations.

The presence of a nonzero feed rate $f$ is rooted in biology.  In general, biological organisms are open, out-of-equilibrium systems with an external input of energy.  As discussed by Morowitz~\cite{Morowitz_1979}, this external energy input is considered to be crucial for the formation of complex structures in living systems. In the case of a vanishing feed rate, all species eventually decay into inert products and the system dies off.  For a nonvanishing feed rate and away from substrate diffusion centers, there exists a constant equilibrium value $U_{\rm eq} = f/r_{u}$ for the substrate.

The system of equations~(\ref{eq:Gray_Scott_equations_1})-(\ref{eq:Gray_Scott_equations_2}) has a conserved quantity in the absence of feed and decay terms.  To see that, note that the diffusion equation is derived from two building blocks: the continuity equation and Fick's law (i.e. the flux of a chemical species is proportional to a gradient of concentration).  Defining the flux of B as ${\bf j}_{b} = -D_{b}\nabla B$, we can rewrite Eqs.~(\ref{eq:Gray_Scott_equations_1})-(\ref{eq:Gray_Scott_equations_2}) as:
\begin{eqnarray}
\label{eq:Continuity_1}
\frac{\partial V}{\partial t} + \nabla\cdot {\bf j}_{v} & = &  - r_{v}V + \lambda UV^{2} + \eta_{v}(x), \\
\label{eq:Continuity_2}
\frac{\partial U}{\partial t} + \nabla\cdot {\bf j}_{u} & = &  - r_{u}U - \lambda UV^{2} + \eta_{u}(x) + f,
\end{eqnarray}
which have the form of a continuity equation.  The terms on the RHS are sources/sinks of U's and V's.  Adding the two equations together and setting $f = r_{u} = r_{v} = 0$ we get:
\begin{eqnarray}
\frac{\partial (V + U)}{\partial t} + \nabla \cdot({\bf j}_{v} + {\bf j}_{u}) & = &  \eta_{v}(x) + \eta_{u}(x).
\end{eqnarray}
Thus the total concentration $U + V$ is conserved on average (if the noise has zero mean).  This conservation law is broken explicitly by the feed and decay terms.  Note that the cancellation of the $\lambda UV^{2}$ terms is a direct consequence of stoichiometry.  This is another way of uncovering the U(1) symmetry found in Ref.~\cite{Cooper_etal_2013} \footnote{The authors of Ref.~\cite{Cooper_etal_2013} analyse the symmetries of the CARD equations from their Hamiltonian.  We do not do that here because it is not possible to write down a Hamiltonian/Lagrangian for diffusion without introducing additional fields (like in the Closed-Time-Path (e.g.~\cite{Calzetta_Hu_2008}) or Martin-Siggia-Rose~\cite{Martin_etal_1973} formalisms). These additional fields are not directly related to physical quantities and make the physical interpretation of results more difficult.}.

Let's rewrite Eqs.~(\ref{eq:Gray_Scott_equations_1})-(\ref{eq:Gray_Scott_equations_2}) in a way more suitable for our purposes.  Define $\tilde{U} \equiv U - U_{\rm eq} = U - f/r_{u}$.  Plugging this definition in the evolution equations~(\ref{eq:Gray_Scott_equations_1})-(\ref{eq:Gray_Scott_equations_2}), we get after simplifications:
\begin{eqnarray}
\label{eq:Gray_Scott_equations_modified_1}
\frac{\partial V}{\partial t} & = & D_{v}\nabla^{2}V - r_{v} V + \lambda \tilde{U} V^{2} + \frac{\lambda f}{r_{u}}V^{2} + \eta_{v}(x), \\
\label{eq:Gray_Scott_equations_modified_2}
\frac{\partial \tilde{U}}{\partial t} & = & D_{u}\nabla^{2}\tilde{U} - r_{u} \tilde{U} - \lambda \tilde{U}V^{2} - \frac{\lambda f}{r_{u}}V^{2} + \eta_{u}(x).
\end{eqnarray}
After this redefinition of the field $U$, the constant feed term disappears.  The price to pay is to have two additional interactions in the evolution equations; they can be treated in the same way as the ``regular'' $\lambda UV^{2}$ interactions, except that their couplings are proportional to the feed rate.

In this paper, we assume that the following condition is satisfied:
\begin{eqnarray}
\label{eq:Approximation_feeding_rate}
\lambda\tilde{U}V^{2} \gg \frac{\lambda f}{r_{u}} V^{2} & \;\;\rightarrow\;\; & U \gg 2\frac{f}{r_{u}},
\end{eqnarray}
i.e. the interactions proportional to the feed rate are negligible compared to the regular interactions.  This happens when the concentration of U is very large compared to its equilibrium value, in which case feeding more U's does not produce significant effects on the amount already present.  A more concrete way of thinking about this approximation goes as follows.  Imagine a Petri dish full of food (U) and a few seeds of organism (V).  The Petri dish is connected to a tube that injects more food into the system at a rate $f$.  The amount of food in the system decreases via decay into inert products or via conversion into organisms.  There exists two limiting initial conditions for the food: $U_{\rm large} \gg U_{\rm eq}$ and $U_{\rm small} \ll U_{\rm eq}$.  For $U_{\rm large}$, there is so much food in the Petri dish that the system evolves by producing organisms without the need of any external feeding.  Since interactions are generally small, the main source of food loss is via exponential decay.  When the initial amount of food at a time $t$ is comparable to the equilibrium value $U_{\rm large}e^{-r_{u}t} \approx U_{\rm eq}$, our approximation breaks down.  This happens for times $t > r_{u}^{-1}\ln(U_{\rm large}/U_{\rm eq})$.  The effect of including the feed term in the renormalization analysis amounts to adding new Feynman diagrams in Fig.~\ref{fig:One_loop_corrections} (due to the additional interaction term $\frac{\lambda f}{r_{u}}V^{2}$).  For the sake of brevity and clarity, we choose to restrict our analysis to the case where there is a lot of food initially in the system and we neglect the feed term (and we drop the tilde over $U$).

\subsection{Role of noise}

An important ingredient in our approach is the presence of noise in Eqs.~(\ref{eq:Gray_Scott_equations_1})-(\ref{eq:Gray_Scott_equations_2}).  Noise adds fluctuations to an otherwise deterministic system and is ubiquitous in physics, chemistry and biology.  Examples include  noise in electrical circuits and solids (e.g.~\cite{MacDonald_2006,Dutta_Horn_1981}), thermal fluctuations due to the random motion of molecules in chemical reactions (e.g.~\cite{MacDonald_2006,Gillespie_2007}), mechanical noise in chemical reactions~\cite{Carnall_etal_2010} and noisy gene expression (e.g.~\cite{Tsimring_2014}).  In equilibrium, fluctuations are usually small due to the law of large numbers.  In an out-of-equilibrium system like the stochastic CARD model, the central limit theorem does not necessarily apply and fluctuations can be larger.

There are two generic types of noise: extrinsic and intrinsic~\cite{vanKampen_2007}.  Extrinsic noise is caused by the application of a random force external to the system.  It includes environmental effects such as thermal fluctuations.  Since no chemical or biological organism is perfectly isolated from the outside world, extrinsic noise is an important ingredient in the analysis of realistic systems.  By contrast, intrinsic noise is still present for a system in complete isolation.  It is caused by the fact that the system itself is made of discrete particles and is inherent in the very mechanism by which the system evolves.  The randomness of quantum mechanical processes is an example of intrinsic noise.  Thus $\eta_{v}$, $\eta_{u}$ contain two components (extrinsic and intrinsic), but there is no way of distinguishing between the two at the level of the partial differential equations.  To make this distinction, an approach based on the Master equation is required~\cite{vanKampen_2007,Cooper_etal_2013,Butler_Goldenfeld_2011}.  This is beyond the scope of this paper.

The deterministic CARD model (i.e. without the noise terms $\eta_{v}$ and $\eta_{u}$ in Eqs.~(\ref{eq:Gray_Scott_equations_1})-(\ref{eq:Gray_Scott_equations_2})) is studied numerically in Ref.~\cite{Pearson_1993}.  The results show the formation of various patterns (incomplete spirals, stripes, self-replicating spots), where their appearance is controlled by the parameters of the model (feed and decay rates).  The stochastic CARD model is studied numerically in Ref.~\cite{Lesmes_etal_2003,Zorzano_etal_2004}.  Although both deterministic and stochastic models produce similar types of patterns, the presence of gaussian white noise affects the formation of patterns in a nontrivial way.  For instance, increasing the noise intensity makes the type of pattern change from growing stripes to spot self-replication.  This is an example of noise-controlled pattern selection.  This last point is crucial for chemical and biological applications, since it establishes a link between the internal structure and environment of a system and what types of patterns are formed.  Conversely, controlled noise in an experiment might be used to probe the microscopic dynamics of a system.  

In the following, we use a two-parameters Gaussian colored noise to describe the stochastic component in Eqs.~(\ref{eq:Gray_Scott_equations_1})-(\ref{eq:Gray_Scott_equations_2}).  Its statistical properties are given by:
\begin{eqnarray}
\label{eq:Noise_property_1}
\langle \eta_{v}(k) \rangle & = & \langle \eta_{u}(k) \rangle \;\;\;=\;\;\; 0, \\
\label{eq:Noise_property_2}
\langle \eta_{v}(k)\eta_{v}(p) \rangle & = & 2A_{v}|{\bf k}|^{-y_{v}}(2\pi)^{d_{s}+1}\delta^{(d_{s}+1)}(k+p), \\
\label{eq:Noise_property_3}
\langle \eta_{u}(k)\eta_{u}(p) \rangle & = & 2A_{u}|{\bf k}|^{-y_{u}}(2\pi)^{d_{s}+1}\delta^{(d_{s}+1)}(k+p), \\
\label{eq:Noise_property_4}
\langle \eta_{v}(k)\eta_{u}(p) \rangle & = & \langle \eta_{u}(k)\eta_{v}(p) \rangle \;\;\;=\;\;\; 0,
\end{eqnarray}
where we use the shortcut notation $k = ({\bf k},\omega)$ and we have expressed the correlations in Fourier space for later convenience.  All higher order moments are zero (Gaussian noise).  Different wavenumbers ${\bf k}$ and frequencies $\omega$ are statistically uncorrelated due to the $\delta^{(d_{s}+1)}$ functions, where $d_{s}$ is the dimension of space.  We restrict ourselves to spatially correlated noise in this paper (see Refs.~\cite{Hochberg_etal_2003,Medina_etal_1989} for examples of temporally correlated noise).  The noise amplitudes $A_{u},A_{v} > 0$ and noise exponents $y_{u},y_{v}$ are free parameters of the model.  The amplitude gives the overall strength of the fluctuations while the exponent gives the strength of correlations as a function of wavenumber.  The model includes Gaussian white noise as a special case ($y_{u},y_{v} = 0$, equal spectral power in all frequencies), red noise ($y_{u},y_{v} > 0$, more spectral power at low frequencies or long distances) and blue noise ($y_{u},y_{v} < 0$, more spectral power at high frequencies or short distances).  The choice of power law noise can be justified in two ways.  First, power laws are found in many natural and man-made systems, and many plausible mechanisms may produce them~\cite{Newman_2005}.  They allow for stronger fluctuations and possibly more complex, scale-dependent patterns.  Second, power laws can be used as a basis to approximate more complex noise functions using Taylor series.  In Sect.~\ref{sec:Renormalization} we show how noise enters field theory computations through loop diagrams and how it affects the parameters of the model.

\section{UV renormalization of the stochastic CARD model \label{sec:Renormalization}}

\subsection{Renormalized equations and counterterms \label{sec:Renormalized_equations}}

To start our renormalization program, we write down the bare stochastic CARD equations in Fourier space:
\begin{eqnarray}
\label{eq:Gray_Scott_equations_Fourier_bare_1}
\lefteqn{(D_{v0}|{\bf k}|^{2} - i\omega + r_{v0})V_{0}} \nonumber \\
 & & \hspace{0.1in} - \lambda_{0}\int d^{d_{s}+1}p_{1} \int d^{d_{s}+1}p_{2}\; U_{0}V_{0}^{2} -\eta_{v0}  =  0, \\
\label{eq:Gray_Scott_equations_Fourier_bare_2}
\lefteqn{(D_{u0}|{\bf k}|^{2} - i\omega + r_{u0})U_{0}} \nonumber \\
 & & \hspace{0.1in}  + \lambda_{0}\int d^{d_{s}+1}p_{1} \int d^{d_{s}+1}p_{2}\; U_{0}V_{0}^{2} -\eta_{u0} - F  =  0,
\end{eqnarray}
where the zero subscript indicates bare parameters, $F = f(2\pi)^{d_{s}+1}\delta(\omega)\delta({\bf k})$ is the Fourier transform of the feed rate and we use the shortcut notation $\int d^{d_{s}+1}p = \int\frac{d^{d_{s}}p}{(2\pi)^{d_{s}}}\int\frac{d\omega}{(2\pi)}$.  The external feed rate $f$ is considered a classical source and does not need to be renormalized.  Loop corrections to the other parameters are computed using the Feynman rules found in Appendix~\ref{sec:Feynman_rules}.  In this paper we limit ourselves to one-loop corrections.  At one-loop order there is no correction to the noises, so we drop the zero subscript on $\eta$'s.  

We use dimensional continuation as our regulator. See Appendix~\ref{sec:Regularization_noise} for a discussion of the subtleties when regularizing in the presence of noise.  Defining the renormalized chemical fields $V_{0} = Z_{v}V$ and $U_{0} = Z_{u}U$, we rewrite the evolution equations in $d$ dimensions as:
\begin{eqnarray}
\label{eq:Gray_Scott_equations_Fourier_renormalized_1}
\lefteqn{\left[(D_{v}+A)|{\bf k}|^{2} - i(1+B)\omega + (r_{v}+C)\right]V} \nonumber \\
 & &  - (\lambda^{(d)} + D)\int d^{d+1}p_{1} \int d^{d+1}p_{2}\; UV^{2} -\eta_{v}  =  0, \\
\label{eq:Gray_Scott_equations_Fourier_renormalized_2}
\lefteqn{\left[(D_{u}+E)|{\bf k}|^{2} - i(1+F)\omega + (r_{u}+G)\right]U} \nonumber \\
 & &   + (\lambda^{(d)}+H)\int d^{d+1}p_{1} \int d^{d+1}p_{2}\; UV^{2} -\eta_{u} - F  =  0,
\end{eqnarray}
where $\lambda^{(d)}$ is the coupling constant in $d$ dimensions and:
\begin{eqnarray}
\begin{array}{ll} A = Z_{v}D_{v0} - D_{v},  &  E = Z_{u}D_{u0} - D_{u}, \\
B = Z_{v} - 1, & F = Z_{u} - 1, \\
C = Z_{v}r_{v0} - r_{v}, & G = Z_{u}r_{u0} - r_{u}, \\
D = Z_{u}Z_{v}^{2}\lambda_{0}^{(d)}- \lambda^{(d)},\;\; & H = Z_{u}Z_{v}^{2}\lambda_{0}^{(d)} - \lambda^{(d)}.
\end{array}
\end{eqnarray}
Parameters without a zero subscript are the renormalized parameters and $A, B, \dots, H$ are counterterms.  

At one-loop order we have ${A=B=C=E=F=0}$, implying $Z_{v} = Z_{u} = 1$, $D_{v0} = D_{v}$, $D_{u0} = D_{u}$ and ${r_{v0} = r_{v}}$.  The only nontrivial one-loop corrections are for $r_{u}$ and $\lambda$ (see Fig.~\ref{fig:One_loop_corrections}).  They are given by (see Appendix~\ref{sec:One_loop_corrections} for details):
\begin{figure}
\includegraphics[width=0.35\textwidth]{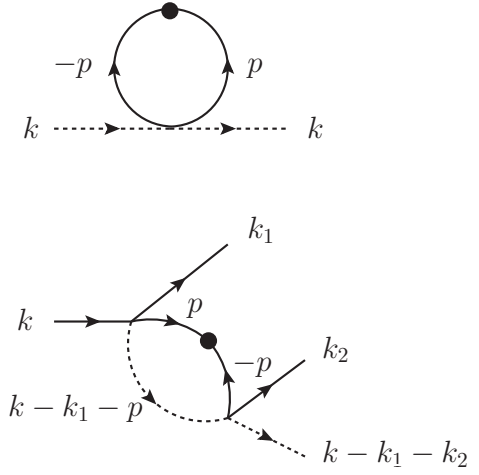}%
\caption{One-loop diagrams contributing to $r_{u}$ and $\lambda$. \label{fig:One_loop_corrections}}
\end{figure}
\begin{eqnarray}
\label{eq:Correction_ru}
\Gamma_{r_{u}} & = & -\frac{\lambda^{(d)}A_{v}^{(d)}}{r_{v}} \left(\frac{r_{v}}{D_{v}}\right)^{\frac{d-y_{v}}{2}} \frac{K_{d}}{2} \frac{\pi}{\sin\left(\pi\frac{(d-y_{v})}{2}\right)}, \\
\label{eq:Correction_lambda}
\Gamma_{\lambda}(0) & = & \frac{4\lambda_{(d)}^{2}A_{v}^{(d)}}{D_{v}(D_{u}+D_{v})}  \frac{K_{d}}{2} \frac{\pi}{\sin\left(\pi\frac{(d-y_{v})}{2}\right)} \nonumber \\
                    &   & \times \left[\frac{\left(\frac{r_{v} + r_{u}}{D_{v} + D_{u}}\right)^{-1+\frac{d-y_{v}}{2}} - \left(\frac{r_{v}}{D_{v}}\right)^{-1+\frac{d-y_{v}}{2}}}{\left(\frac{r_{v} + r_{u}}{D_{v} + D_{u}}\right) - \left(\frac{r_{v}}{D_{v}}\right)}\right],
\end{eqnarray}
where we use the shortcut notation $K_{d} = 2/(4\pi)^{\frac{d}{2}}\Gamma\left(\frac{d}{2}\right)$.  Note that we take all external momenta to be zero (hydrodynamic limit) in the above one-loop computations.  This is sufficient for beta function computations of marginal operator couplings, since external momenta corrections are confined to logarithmic finite terms in this case.  

The sine functions in Eqs.~(\ref{eq:Correction_ru})-(\ref{eq:Correction_lambda}) produce poles at ${d-y_{v} = 2m}$ with $m \in {\mathbb{Z}}$.  To extract the poles, we define $d_{m} = y_{v} + 2m$ and expand around $d = d_{m} - \epsilon$.  At leading order in $\epsilon$ we obtain:
\begin{eqnarray}
\label{eq:Correction_ru_pole}
\Gamma_{r_{u}} & = & -\frac{\lambda^{(d_{m})}A_{v}^{(d_{m})}}{r_{v}} \left(\frac{r_{v}}{D_{v}}\right)^{m} (-1)^{m+1} K_{d} \;\;\frac{1}{\epsilon},  \\
\label{eq:Correction_lambda_pole}
\Gamma_{\lambda}(0) & = & 4\lambda^{(d)}\frac{\lambda^{(d_{m})}A_{v}^{(d_{m})}}{D_{v}(D_{u}+D_{v})} (-1)^{m+1} K_{d} \;\;\frac{1}{\epsilon} \nonumber \\
                    &   & \times \left[\frac{\left(\frac{r_{v} + r_{u}}{D_{v} + D_{u}}\right)^{-1+m} - \left(\frac{r_{v}}{D_{v}}\right)^{-1+m}}{\left(\frac{r_{v} + r_{u}}{D_{v} + D_{u}}\right) - \left(\frac{r_{v}}{D_{v}}\right)}\right].
\end{eqnarray}
We use the MS prescription to define counterterms, implying $D = H = \Gamma_{\lambda}(0)$ and $G = \Gamma_{r_{u}}(0)$.  The corresponding $Z$ factors are given by:
\begin{eqnarray}
\label{eq:Z_factor_ru}
r_{u0} & = &  \left(1 + \frac{\Gamma_{r_{u}}(0)}{r_{u}}\right)r_{u} \;\;\equiv\;\; Z_{r_{u}}r_{u}, \\
\label{eq:Z_factor_lambda}
\lambda_{0}^{(d)} & = & \left(1 + \frac{\Gamma_{\lambda}(0)}{\lambda^{(d)}}\right)\lambda^{(d)} \;\;\equiv\;\; Z_{\lambda}\lambda^{(d)}.
\end{eqnarray}
%

\subsection{Beta function computations \label{sec:Beta_function_computations}}

\begin{figure}
\includegraphics[width=0.45\textwidth]{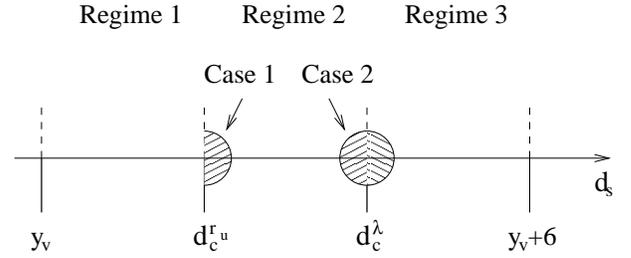}
\caption{Schematic representation of the three possible regimes for beta function computations.  The shaded half circles represent the range of validity of the different $\epsilon$-expansions we use.  
\label{fig:General_scheme}}
\end{figure}

\begin{figure}
\includegraphics[width=0.25\textwidth]{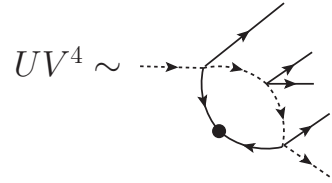}
\caption{Example of one-loop higher order operator.  This diagram is UV-divergent for $d _{s} \geq y_{v} + 6$ and requires a new counterterm not present at tree-level.
\label{fig:Higher_order_operator}}
\end{figure}

As explained in Appendix~\ref{sec:Regularization_noise}, the noise exponents $y_{b}$ dictate the divergence structure of loop integrals.  Different parameters have different critical dimensions $d_{c}$ at which they start to require renormalization.  For instance we have (c.f. Eqs.~(\ref{eq:Correction_ru_after_frequency}) and (\ref{eq:Correction_after_frequency})):
\begin{eqnarray}
\Gamma_{a} \sim A_{v} \int d|{\bf p}|\;|{\bf p}|^{d_{s}-d_{c}^{a}-1},
\end{eqnarray}
with $d_{c}^{r_{u}} = y_{v} + 2$ and $d_{c}^{\lambda} = y_{v} + 4$.  We can identify three regimes (see Fig.~\ref{fig:General_scheme}).  For $y_{v} < d_{s} < d_{c}^{r_{u}}$ (regime 1), both one-loop corrections are finite and no renormalization is required.  For $d_{c}^{r_{u}} \leq d_{s} < d_{c}^{\lambda}$ (regime 2), only $r_{u}$ diverges and requires renormalization.  For $d_{c}^{\lambda} \leq d_{s} < y_{v} + 6$ (regime 3), both $r_{u}$ and $\lambda$ have to be renormalized.  The case $d_{s} \leq y_{v}$ corresponds to noise-induced IR divergences and we leave them for future work.  For $d_{s} \geq y_{v} + 6$, new UV-divergent operators not present at tree-level are generated.  An example is shown in Fig.~\ref{fig:Higher_order_operator}.  A full discussion of those noise-induced higher order operators goes beyond the aim and scope of the present paper.  It is shown in Ref.~\cite{Gagnon_PerezMercader_2015} that those higher order operators are nonrenormalizable and are suppressed at low energies.  We leave them out of the present discussion and consider them in future work.

We use the general one-loop results~(\ref{eq:Correction_ru_pole})-(\ref{eq:Correction_lambda_pole}) to compute the beta functions in each regime.  We do an $\epsilon$-expansion around the critical dimensions using $d = d_{c}^{a} - \epsilon$.  The sign of $\epsilon$ tells on which side of the critical dimension we are expanding.  The magnitude of $\epsilon$ tells how far from the critical dimension the system is.  Note that, contrary to particle physics, we do not need to take the $\epsilon\rightarrow 0$ limit at the end of the computation.  This is because the noise exponent $y_{v}$ is an external parameter and can take any value, even fractional ones.  Said differently, power law noise enables control over the effective dimension of a system, and consequently over its divergence structure.

\paragraph*{Regime 1:}  In this regime both factors $Z_{r_{u}}$ and $Z_{\lambda}$ are one.  Thus the scaling of both beta functions $\beta_{r_{u}}$ and $\beta_{\lambda}$ is trivial.

\paragraph*{Regime 2:} In this regime only $\Gamma_{r_{u}}$ diverges, thus the scaling of $\beta_{\lambda}$ is trivial.  There are two ways to compute $\beta_{r_{u}}$ in this regime: either expand around $d = d_{c}^{r_{u}} + |\epsilon|$ (case 1) or around $d = d_{c}^{\lambda} - |\epsilon|$ (case 2) (see Fig.~\ref{fig:General_scheme}).  We study both cases separately.

The Z-factors for case 1 are obtained from Eqs.~(\ref{eq:Correction_ru_pole})-(\ref{eq:Z_factor_lambda}) with $m=1$:
\begin{eqnarray}
Z_{r_{u}} & = &  1 + g^{(d)}L^{-|\epsilon|} \frac{K_{d}}{|\epsilon|}, \\
Z_{\lambda} & = & 1,
\end{eqnarray}
where $L$ is an arbitrary length scale and
\begin{equation}
g^{(d)}  \equiv  \frac{\lambda^{(d)}A_{v}^{(d)}}{r_{u} D_{v}} \;\;=\;\; \frac{\lambda^{(d_{c}^{r_{u}})}A_{v}^{(d_{c}^{r_{u}})}}{r_{u} D_{v}} L^{|\epsilon|} \;\;\equiv\;\; g^{(d_{c}^{r_{u}})}L^{|\epsilon|},
\end{equation}
is an effective coupling with vanishing engineering dimension when $\epsilon\rightarrow 0$.  The relation between the bare and renormalized effective coupling is:
\begin{equation}
\label{eq:Effective_coupling_regime1_case1_g}
g_{0}^{(d)} = \frac{\lambda_{0}^{(d)}A_{v}^{(d)}}{r_{u0}D_{v}} \;\;=\;\; \frac{Z_{\lambda}}{Z_{r_{u}}} \frac{\lambda^{(d)}A_{v}^{(d)}}{r_{u} D_{v}} \;\;=\;\; \frac{Z_{\lambda}}{Z_{r_{u}}} g^{(d)},
\end{equation}
Differentiating on both sides with respect to the arbitrary length scale $L$ and simplifying, we obtain the beta function for the effective coupling:
\begin{equation}
\label{eq:Beta_fct_regime2_case1_g}
\beta_{g}  \equiv  L\frac{d g^{(d_{c}^{r_{u}})}}{dL} \;\;=\;\; g^{(d_{c}^{r_{u}})}\left(-|\epsilon| - g^{(d_{c}^{r_{u}})}K_{d}\right)  + O(g^{3}).
\end{equation}
The above result is valid for $g^{(d_{c}^{r_{u}})} < 1$.  To interpret physically, we revert back to the original parameters with a change of variables.  This gives:
\begin{equation}
\label{eq:Beta_fct_regime2_case1}
\beta_{r_{u}} \;\;\equiv\;\; L\frac{d r_{u}}{dL} \;\;=\;\; |\epsilon|r_{u} + \frac{\lambda^{(d_{c}^{r_{u}})} A_{v}}{D_{v}}K_{d} + O(\lambda^{2}).
\end{equation}
The beta function for case 2 is obtained in a similar way, using Eqs.~(\ref{eq:Correction_ru_pole}) and (\ref{eq:Z_factor_ru}) with $m=2$.  The result is:
\begin{eqnarray}
\label{eq:Beta_fct_regime2_case2_g}
\beta_{g} & = & g^{(d_{c}^{\lambda})}\left(|\epsilon| + g^{(d_{c}^{\lambda})}K_{d}\right) + O(g^{3}),
\end{eqnarray}
or, in terms of the original parameters:
\begin{eqnarray}
\label{eq:Beta_fct_regime2_case2}
\beta_{r_{u}} & = & -\left(|\epsilon|r_{u} + \frac{\lambda A_{v}r_{v}}{D_{v}^{2}}K_{d}\right) + O(\lambda^{2}),
\end{eqnarray}
where the effective coupling is given by:
\begin{eqnarray}
\label{eq:Effective_coupling_regime1_case2_g}
g^{(d)} & = & \frac{\lambda^{(d)}A_{v}^{(d)}r_{v}}{r_{u} D_{v}^{2}} \;\;=\;\; \frac{\lambda^{(d_{c}^{\lambda})}A_{v}^{(d_{c}^{\lambda})}r_{v}}{r_{u} D_{v}^{2}} L^{-|\epsilon|}.
\end{eqnarray}

\paragraph*{Regime 3:} In this regime both $\Gamma_{r_{u}}$ and $\Gamma_{\lambda}$ diverge, resulting in two nontrivial beta functions.  To compute them, we can either expand around $d = d_{c}^{\lambda} + |\epsilon|$ or $d = (y_{v} + 6) - |\epsilon|$.  For the sake of brevity, we only expand around the former.

The Z-factors are obtained from Eqs.~(\ref{eq:Correction_ru_pole})-(\ref{eq:Z_factor_lambda}) with $m=2$:
\begin{eqnarray}
Z_{r_{u}} & = & 1 - g^{(d)}L^{-|\epsilon|} \frac{K_{d}}{|\epsilon|}, \\
Z_{\lambda} & = & 1 + 4h^{(d)}L^{-|\epsilon|}\frac{K_{d}}{|\epsilon|},
\end{eqnarray}
where:
\begin{eqnarray}
\label{eq:Effective_coupling_regime3_g}
g^{(d)} & = & \frac{\lambda^{(d)}A_{v}^{(d)}r_{v}}{r_{u} D_{v}^{2}} \;\;=\;\; \frac{\lambda^{(d_{c}^{\lambda})}A_{v}^{(d_{c}^{\lambda})}r_{v}}{r_{u} D_{v}^{2}} L^{|\epsilon|}, \\
\label{eq:Effective_coupling_regime3_h}
h^{(d)} & = & \frac{\lambda^{(d)}A_{v}^{(d)}}{D_{v}(D_{u} + D_{v})} \;\;=\;\;  \frac{\lambda^{(d_{c}^{\lambda})}A_{v}^{(d_{c}^{\lambda})}}{D_{v}(D_{u} + D_{v})} L^{|\epsilon|}.
\end{eqnarray}
Both effective couplings are dimensionless when $\epsilon\rightarrow 0$.  The relations between bare and renormalized effective couplings are:
\begin{eqnarray}
g_{0}^{(d)} & = & \frac{\lambda_{0}^{(d)}A_{v}^{(d)}r_{v}}{r_{u0} D_{v}^{2}}  \;\;=\;\; \frac{Z_{\lambda}}{Z_{r_{u}}}g^{(d)}, \\
h_{0}^{(d)}  & = & \frac{\lambda_{0}^{(d)}A_{v}^{(d)}}{D_{v}(D_{u} + D_{v})}  \;\;=\;\; Z_{\lambda} h^{(d)}.
\end{eqnarray}
Differentiating on both sides with respect to the arbitrary length scale $L$ and simplifying, we get:
\begin{eqnarray}
\beta_{g} & = & g^{(d_{c}^{\lambda})}\left(-|\epsilon| + g^{(d_{c}^{\lambda})}K_{d} + 4h^{(d_{c}^{\lambda})}K_{d}\right) + O(g^{3},h^{3}), \\
\beta_{h} & = & h^{(d_{c}^{\lambda})}\left(-|\epsilon| +4h^{(d_{c}^{\lambda})}K_{d}\right) + O(g^{3},h^{3}).
\end{eqnarray}
Reverting back to the original parameters, we get:
\begin{eqnarray}
\label{eq:Beta_fct_regime3_ru}
\beta_{r_{u}} & = & -\frac{\lambda A_{v}r_{v}}{D_{v}^{2}}K_{d} + O(\lambda^{3}), \\
\label{eq:Beta_fct_regime3_lambda}
\beta_{\lambda} & = & -|\epsilon|\lambda + \frac{4A_{v}}{D_{v}(D_{u}+D_{v})}\lambda^{2}K_{d} + O(\lambda^{3}).
\end{eqnarray}
We analyze and discuss the obtained beta functions in Sect.~\ref{sec:Discussion}.

\section{Results and discussion \label{sec:Discussion}}

In this section, we study the behavior of the beta functions obtained for regimes 2 and 3 (there is no running of parameters in regime 1).  The corresponding running solutions are also obtained and interpreted.  Note that the plots in this section are representative examples.  We use the same parameters for all plots in order to make comparisons easier.  Since the CARD model has no direct analog in real chemistry, an exhaustive study of the parameter space is not necessary.  Our goal is to point out interesting qualitative behaviors arising from power-law fluctuations that may be present in similar real chemical systems.

\begin{figure}
\includegraphics[width=0.5\textwidth]{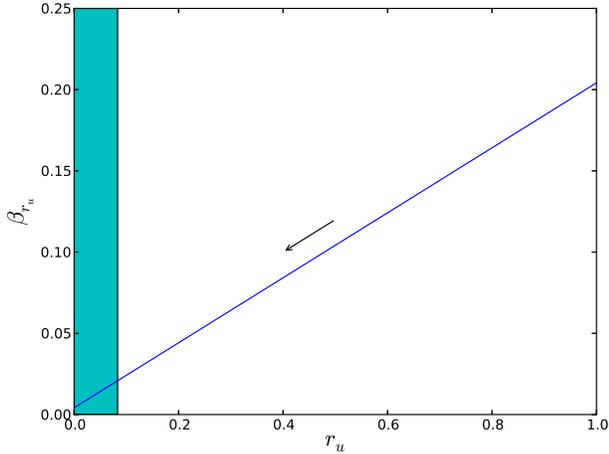}%
\caption{Renormalization group flow diagram for case 1 of regime 2.  We used $|\epsilon| = 0.2$, $K_{3} = 0.05$, $A_{v} = 0.5$, $D_{v} = 0.3$, $\lambda = 0.05$ for the plotting.  There is no fixed point anywhere.  Shaded region indicates breakdown of perturbation theory (i.e. $g>1$ for $r_{u} < 0.08$).  Arrow indicates direction of decreasing $L$. \label{fig:Beta_function_regime2_case1}}
\end{figure}

\paragraph*{Regime 2, case 1:} The beta function~(\ref{eq:Beta_fct_regime2_case1}) has no fixed point and its derivative is $d\beta_{r_{u}}/dr_{u} = |\epsilon| > 0$.  It is thus always positive and increases monotonically.  A plot of the beta function for this case is shown in Fig.~\ref{fig:Beta_function_regime2_case1}.

\begin{figure}
\includegraphics[width=0.5\textwidth]{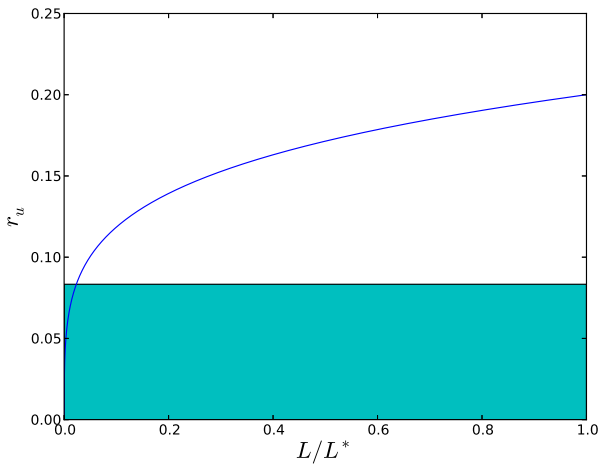}%
\caption{Plot of the running decay rate for case 1 of regime 2.  We used $|\epsilon| = 0.2$, $K_{d} = 0.05$, $A_{v} = 0.5$, $D_{v} = 0.3$, $\lambda = 0.05$, $r_{u}(L^{*}) = 0.2$ for the plotting.  Shaded region indicates breakdown of perturbation theory (i.e. $g>1$ for $r_{u} < 0.08$).  \label{fig:Plot_nu_regime2_case1}}
\end{figure}

The corresponding solution for the running decay rate is obtained by integrating Eq.~(\ref{eq:Beta_fct_regime2_case1}).  The result is:
\begin{equation}
\label{eq:Plot_nu_regime1_case1}
r_{u}(L)  =  \left(r_{u}(L^{*})+ \frac{\lambda A_{v}K_{d}}{|\epsilon|D_{v}}\right)\left(\frac{L}{L^{*}}\right)^{|\epsilon|} - \frac{\lambda A_{v}K_{d}}{|\epsilon| D_{v}},
\end{equation}
where $L^{*}$ is some scale at which $r_{u}(L^{*})$ is known and can be measured.  A plot of $r_{u}(L)$ is shown in Fig.~\ref{fig:Plot_nu_regime2_case1}.

Note that $\beta_{g}$ (c.f. Eq.~(\ref{eq:Beta_fct_regime2_case1_g})) is obtained as an expansion in the effective coupling $g$ and is thus only valid for $g < 1$.  Using Eq.~(\ref{eq:Effective_coupling_regime1_case1_g}), it translates into a perturbative regime for Eqs.~(\ref{eq:Beta_fct_regime2_case1}) and~(\ref{eq:Plot_nu_regime1_case1}) given by $r_{u} > \lambda A_{v}/D_{v}$.

From Eq.~(\ref{eq:Plot_nu_regime1_case1}) and Fig.~\ref{fig:Plot_nu_regime2_case1}, we infer that the decay rate $r_{u}$ decreases with scale when the dimension of the system is just above $d_{c}^{r_{u}}$.  The decay rate becomes negative for small values of $L$, but this behavior always lies in the region where perturbation theory cannot be trusted.  

In the living system interpretation, it means that the presence of environmental fluctuations in the food+organism system decreases the rate at which food is removed from the system at small scales.  Since the growth of an organism (i.e. $\partial V/\partial t$) is proportional to the amount of food present (i.e. $\lambda UV^{2}$), the growth of structures at smaller scales is increased by fluctuations.

\begin{figure}
\includegraphics[width=0.5\textwidth]{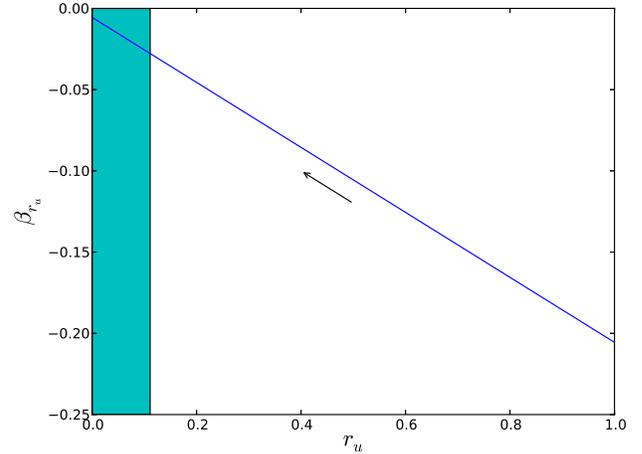}%
\caption{Renormalization group flow diagram for case 2 of regime 2.  We used $|\epsilon| = 0.2$, $K_{d} = 0.05$, $A_{v} = 0.5$, $D_{v} = 0.3$, $r_{v} = 0.4$, $\lambda = 0.05$ for the plotting.  There is no fixed point anywhere.  Shaded region indicates breakdown of perturbation theory (i.e. $g>1$ for $r_{u} < 0.11$).  Arrow indicates direction of decreasing $L$. \label{fig:Beta_function_regime2_case2}}
\end{figure}

\paragraph*{Regime 2, case 2:} The beta function~(\ref{eq:Beta_fct_regime2_case2}) has no fixed point and its derivative is $d\beta_{r_{u}}/dr_{u} = -|\epsilon| < 0$.  It is thus always negative and decreases monotonically.  The beta function is plotted in Fig.~\ref{fig:Beta_function_regime2_case2}.  The corresponding running decay rate is:
\begin{equation}
\label{eq:Plot_nu_regime1_case2}
r_{u}(L)  =  \left(r_{u}(L^{*}) + \frac{\lambda A_{v}r_{v}K_{d}}{|\epsilon|D_{v}^{2}}\right)\left(\frac{L}{L^{*}}\right)^{-|\epsilon|} - \frac{\lambda A_{v}r_{v} K_{d}}{|\epsilon|D_{v}^{2}}.
\end{equation}   
The perturbative regime for Eqs.~(\ref{eq:Beta_fct_regime2_case2}) and (\ref{eq:Plot_nu_regime1_case2}) lies in the region where $r_{u} > \lambda A_{v} r_{v}/D_{v}^{2}$.

\begin{figure}
\includegraphics[width=0.5\textwidth]{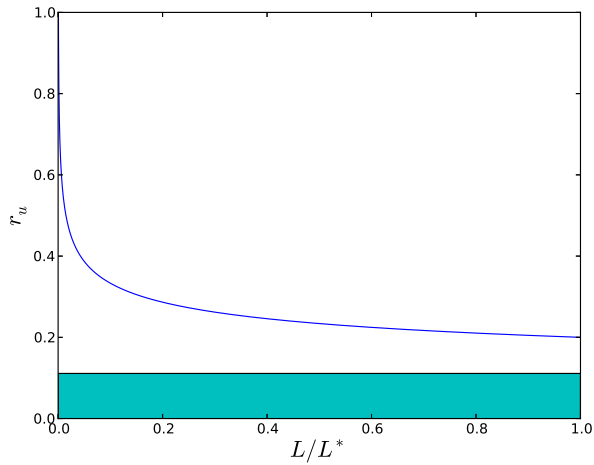}%
\caption{Plot of the running decay rate for case 2 of regime 2.  We used $|\epsilon| = 0.2$, $K_{d} = 0.05$, $A_{v} = 0.5$, $D_{v} = 0.3$, $r_{v} = 0.4$, $\lambda = 0.05$, $r_{u}(L^{*}) = 0.2$ for the plotting.  Shaded region indicates breakdown of perturbation theory (i.e. $g>1$ for $r_{u} < 0.11$).  \label{fig:Plot_nu_regime2_case2}}
\end{figure}

From Eq.~(\ref{eq:Plot_nu_regime1_case2}) and Fig.~\ref{fig:Plot_nu_regime2_case2}, we infer that the decay rate $r_{u}$ increases with scale when the dimension of the system is just below $d_{c}^{\lambda}$.  The decay rate diverges in a power law fashion at $L = 0$.  This behavior is the opposite of the power law decrease of the decay rate just above $d_{c}^{r_{u}}$.  In the living system interpretation, this implies that the growth of structures is dampened at smaller scales.

This drastic change of behavior in the decay rate $r_{u}$ depends on the noise exponent $y_{v}$.  We know that $r_{u}$ gets renormalized (regime 2) when the dimension of space $d_{s}$ lies between $y_{v} + 2 < d_{s} < y_{v} + 4$ (c.f. Sec.~\ref{sec:Beta_function_computations}).  Thus by varying $y_{v}$, it is possible to go from one end of the regime to the other.  This seems to be a general feature for  any parameters that can ultimately be traced to the $(-1)^{m+1}$ factor appearing in the one-loop expressions~(\ref{eq:Correction_ru_pole})-(\ref{eq:Correction_lambda_pole}).  In particular, a drastic change in the properties of a coupling (by varying the noise) could lead to a change in chemical pathway.  This is very similar to the situation in Ref.~\cite{Carnall_etal_2010}, where the authors induce a change in chemical pathway by switching from stirring to shaking in a noise-controlled experiment.

\begin{figure}
\includegraphics[width=0.50\textwidth]{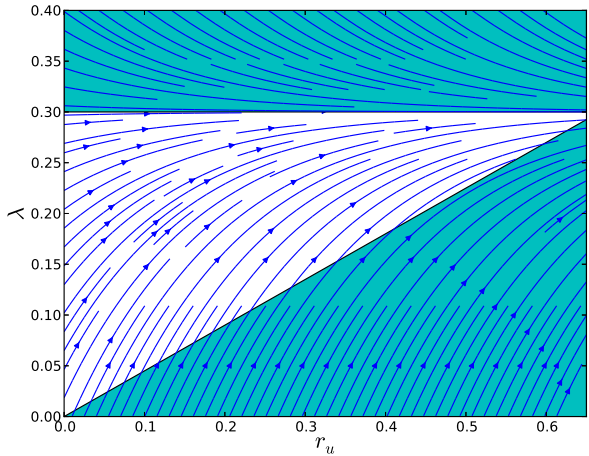}%
\caption{Renormalization group flow diagram for regime 3.  We used $|\epsilon| = 0.2$, $K_{d} = 0.05$, $A_{v} = 0.5$, $D_{v} = 0.3$, $D_{u} = 0.2$, $r_{v} = 0.4$ for the plotting.  There is a line of non-isolated unstable fixed points at $\lambda=0$.  Shaded regions indicate breakdown of perturbation theory (i.e. $g > 1$ for $r_{u} < 2.2\lambda$ and $h > 1$ for $\lambda>0.3$).  Arrows indicate direction of decreasing $L$. \label{fig:Beta_function_regime3}}
\end{figure}

\begin{figure}
\includegraphics[width=0.5\textwidth]{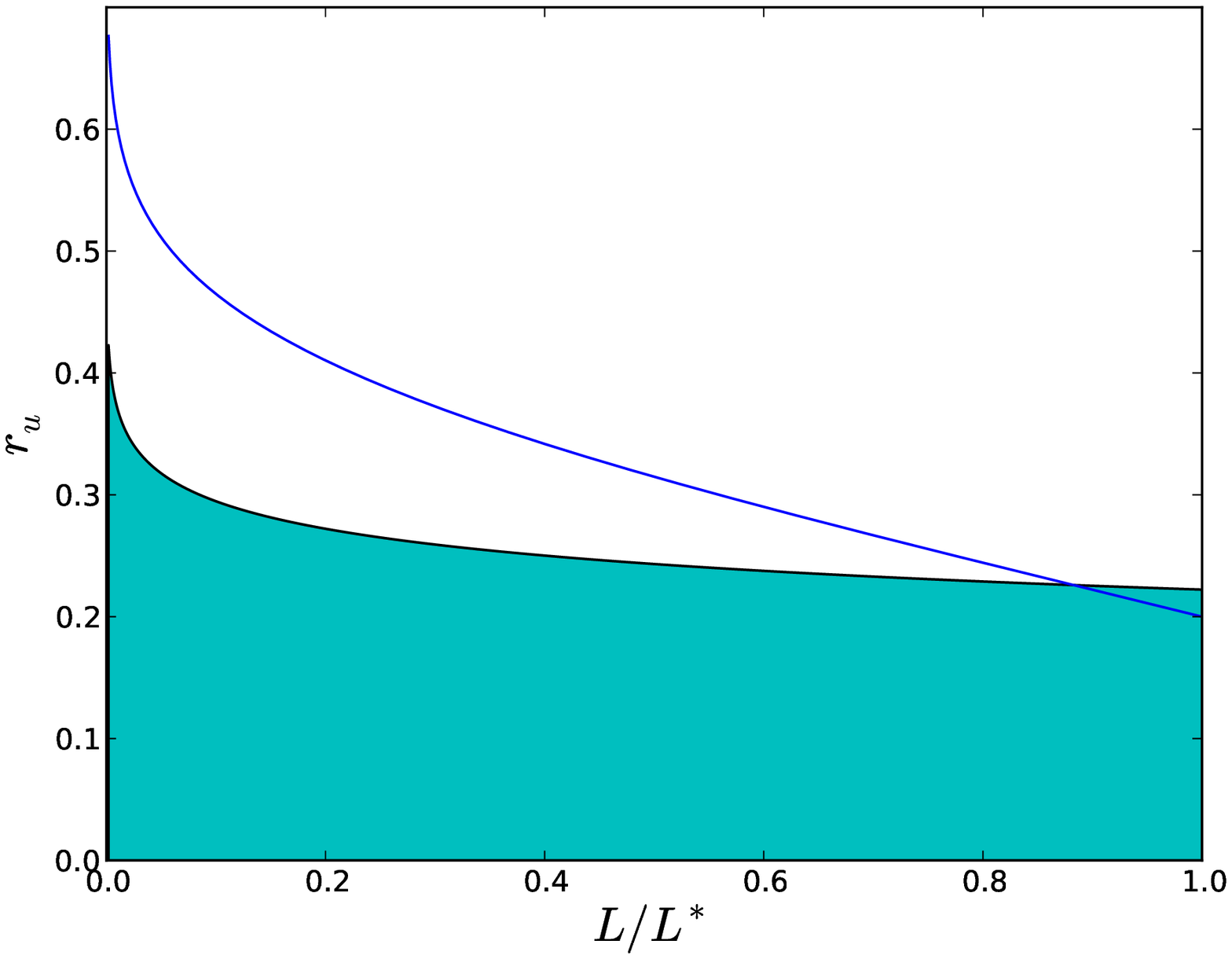}%
\caption{Plot of the running decay rate for regime 3.  We used $|\epsilon| = 0.2$, $K_{d} = 0.05$, $A_{v} = 0.5$, $D_{v} = 0.3$, $D_{u} = 0.2$, $r_{v} = 0.4$, $r_{u}(L^{*}) = 0.2$, $\lambda(L^{*}) = 0.1$ for the plotting.  Shaded region indicates breakdown of perturbation theory (i.e. $g > 1$ for $r_{u} < 2.2\lambda$).  \label{fig:Plot_nu_regime3}}
\end{figure}

\begin{figure}
\includegraphics[width=0.5\textwidth]{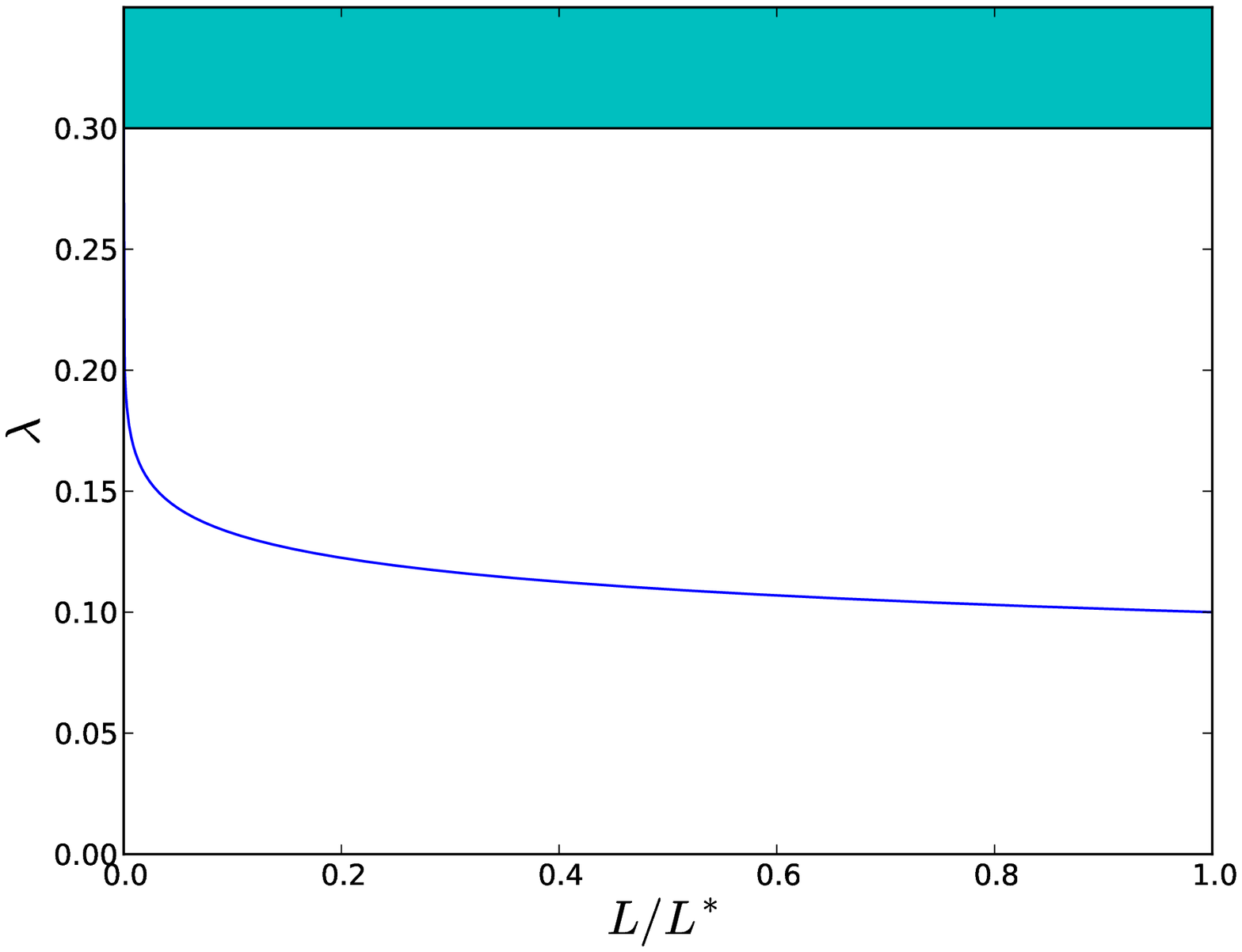}
\caption{Plot of the running coupling for regime 3.  We used $|\epsilon| = 0.2$, $K_{d} = 0.05$, $A_{v} = 0.5$, $D_{v} = 0.3$, $D_{u} = 0.2$, $\lambda(L^{*}) = 0.1$ for the plotting.  Shaded region indicates breakdown of perturbation theory (i.e. $h > 1$ for $\lambda > 0.3$).  \label{fig:Plot_lambda_regime3_1}}
\end{figure}

\begin{figure}
\includegraphics[width=0.5\textwidth]{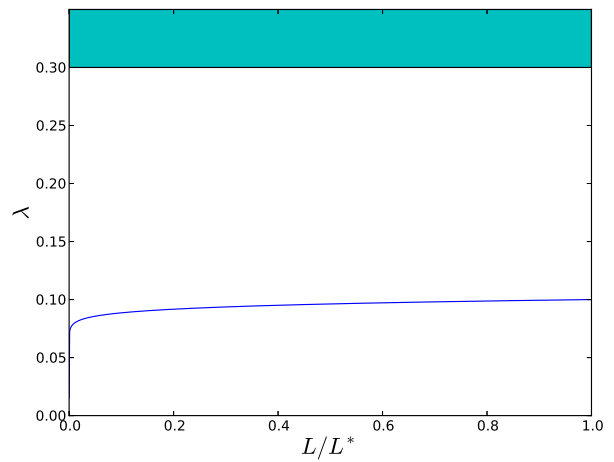}
\caption{Plot of the running coupling for regime 3.  We used $|\epsilon| = 0.01$, $K_{d} = 0.05$, $A_{v} = 0.5$, $D_{v} = 0.3$, $D_{u} = 0.2$, $\lambda(L^{*}) = 0.1$ for the plotting.  Shaded region indicates breakdown of perturbation theory (i.e. $h > 1$ for $\lambda > 0.3$).  \label{fig:Plot_lambda_regime3_2}}
\end{figure}

\paragraph*{Regime 3:} The beta functions~(\ref{eq:Beta_fct_regime3_ru}) and (\ref{eq:Beta_fct_regime3_lambda}) have a line of non-isolated unstable fixed points at $\lambda = 0$.  We also note that $\beta_{\lambda} = 0$ while $\beta_{r_{u}} \neq 0$ at the nontrivial value $\lambda = |\epsilon|D_{v}(D_{u}+D_{v})/4A_{v}K_{d}$.  The renormalization group flow for this regime is shown in Fig.~\ref{fig:Beta_function_regime3}.  The solutions for the decay rate and coupling are:
\begin{eqnarray}
\label{eq:Plot_nu_regime3_ru}
r_{u}(L) & = & r_{u}(L^{*}) + \frac{r_{v}(D_{u}+D_{v})}{4D_{v}} \nonumber \\
         &   & \times \ln\left[1 + \frac{4A_{v}\lambda(L^{*})}{|\epsilon|D_{v}(D_{u}+D_{v})}\left(\left(\frac{L}{L^{*}}\right)^{-|\epsilon|} -1 \right)   \right], \nonumber \\
         &   & \\
\label{eq:Plot_nu_regime3_lambda}
\lambda(L) & = & \frac{|\epsilon|D_{v}(D_{u}+D_{v})}{4A_{v}K_{d}}\frac{1}{1 - \left(1 - \frac{|\epsilon|D_{v}(D_{u}+D_{v})}{4A_{v}K_{d}\lambda(L^{*})}\right) \left(\frac{L}{L^{*}}\right)^{|\epsilon|}}, \nonumber \\
\end{eqnarray}
where $r_{u}(L^{*})$ and $\lambda(L^{*})$ are integration constants fixed by renormalization conditions.  Plots of the two functions are shown in Figs.~\ref{fig:Plot_nu_regime3}-\ref{fig:Plot_lambda_regime3_2}.  The perturbative regime for Eqs.~(\ref{eq:Beta_fct_regime3_ru})-(\ref{eq:Beta_fct_regime3_lambda}) and (\ref{eq:Plot_nu_regime3_ru})-(\ref{eq:Plot_nu_regime3_lambda}) is given by $r_{u} > \lambda A_{v}r_{v}/D_{v}^{2}$ and $\lambda < D_{v}(D_{u}+D_{v})/A_{v}$.

The behavior of the decay rate for regime 3 is similar to the one for regime 2 (case 2), namely the growth of structures at smaller scales is dampened by fluctuations.  There is also the possibility that $r_{u}$ could be negative or develop an imaginary part due to the logarithmic dependence on $L$, but it can be shown that those behaviors lie outside of the domain of validity of perturbation theory.

The coupling has two different types of behavior, depending on the parameters.  For $|\epsilon|D_{v}(D_{u}+D_{v})/4A_{v}K_{d}\lambda(L^{*}) > 1$, the coupling increases at smaller scales and for $|\epsilon|D_{v}(D_{u}+D_{v})/4A_{v}K_{d}\lambda(L^{*}) < 1$, the coupling decreases at smaller scales (compare Figs~\ref{fig:Plot_lambda_regime3_1} and \ref{fig:Plot_lambda_regime3_2}).  In the exceptional case where $|\epsilon|D_{v}(D_{u}+D_{v})/4A_{v}K_{d}\lambda(L^{*}) = 1$, the coupling has no scale dependence at all.  In all cases, it has a finite value at $L=0$ given by $|\epsilon|D_{v}(D_{u}+D_{v})/4A_{v}K_{d}$, which always lies in the perturbative regime.  The model is thus not trivial in the UV.  We note also that the coupling diverges at $L/L^{*} = \left|1 - |\epsilon|D_{v}(D_{u}+D_{v})/4A_{v}K_{d}\lambda(L^{*}) \right|^{-1/|\epsilon|}$, but that this divergence always lies in the region where perturbation theory is not valid.

From the above analysis, we see that noise can have an important influence on the dynamics of a chemical system at small scales.  For instance, for the parameter values specified in Figs.~\ref{fig:Plot_lambda_regime3_1}-\ref{fig:Plot_lambda_regime3_2} we have $|\epsilon|D_{v}(D_{u}+D_{v})/4A_{v}K_{d}\lambda(L^{*}) = 15|\epsilon|$.  This particular combination of parameters is equal to one when $|\epsilon| = 0.067$.  For $|\epsilon|$ greater (less) than 0.067, the coupling increases (decreases) at smaller scales (compare Figs.~\ref{fig:Plot_lambda_regime3_1} and \ref{fig:Plot_lambda_regime3_2}).  Thus varying the noise exponent directly affects the strength of a chemical reaction and might lead to a different chemical pathway.  This is again similar to the situation in Ref.~\cite{Carnall_etal_2010}.

In the living system interpretation of the CARD model, the presence of fluctuations in the environment has two effects.  On the one hand, it increases the rate at which food is removed from the system, leading to a diminished growth of structures at smaller scales.  On the other hand, this effect is either partially compensated or amplified by the change in the coupling $\lambda$ at smaller scales.  For fixed parameter values, the change in coupling is dictated by the noise exponent.

From the above considerations, we see how correlations in environmental noise can directly affect the behavior of the CARD model.  For example, a system in two spatial dimensions subject to gaussian white noise (corresponding to $d_{s}-y_{b} = 2$, regime 2 case 1) has only one running parameter $r_{u}$, while the same system subjected to spatially correlated noise with $y_{b} = -2$ (corresponding to $d_{s}-y_{b} = 4$, regime 3) would exhibit two running parameters $r_{u}$ and $\lambda$.  Those two very different qualitative behaviors are directly related to the value of the noise exponent $y_{b}$.

\section{Conclusion \label{sec:Conclusion}}

In this paper, we study the effect of power law noise on the small scale properties of a cubic autocatalytic reaction-diffusion system (used as a proxy for a very simplified living system).  We show explicitly how noise influences the growth of structures at smaller scales by changing the value of the decay rate and coupling.

The change in parameters depends crucially on the divergence structure of the equations, itself dictated by the noise exponent.  This direct link between external noise and structures at small scales leads to the idea of using noise as a probe to study chemical reactions, in analogy to the way beam energy is used to probe properties of matter in nuclear or particle physics.  

Future work along those lines include working beyond the approximation of large $U$ compared to the feeding rate (c.f. Eq. (\ref{eq:Approximation_feeding_rate})) in order to make contact with more realistic experimental setups.  The extension to temporally correlated noise (in addition to the spatially correlated noise case considered in this paper) is also an important step to make contact with experiments.  To properly implement the fine-graining program advocated in the Introduction, a systematic study of higher order operators is also necessary.  This is work in progress.

\appendix

\section{Regularization in the presence of noise \label{sec:Regularization_noise}}

Before computing the beta functions proper, we discuss some subtleties that arise when regularizing one-loop diagrams in the presence of noise.  The most common regulator in high energy physics is dimensional regularization, because it is very convenient and, more importantly, it preserves gauge symmetry.  There is no gauge symmetry in the CARD model, so using dimensional regularization is not mandatory.  We still use it for all our computations in this paper because of its convenience, but its results must be interpreted carefully.  To help interpreting the results, we use both dimensional and momentum regularization on a generic one-loop integral and show that they give the same results (with some caveats).

\subsection{Dimensional regularization \label{sec:Dimensional_regularization}}

The type of integrals relevant for one-loop computations are of the form:
\begin{eqnarray}
\label{eq:Generic_integral}
I(n,d_{s},y) & = & \int \frac{d^{d_{s}}p}{(2\pi)^{d_{s}}}\; |{\bf p}|^{-y}\frac{1}{(|{\bf p}|^{2} + \Delta^{2})^{n}}, \nonumber \\
\label{eq:Integral_starting_point}
	   & = & \int\frac{d^{d_{s}}\Omega}{(2\pi)^{d_{s}}}\int_{0}^{\infty} d|{\bf p}|\; \frac{|{\bf p}|^{d_{s}-1-y}}{(|{\bf p}|^{2} + \Delta^{2})^{n}},
\end{eqnarray}
where $d_{s}$ is the dimension of space and $y$ is the noise exponent parameter.  The momentum integral~(\ref{eq:Generic_integral}) may diverge depending on the values of $n$, $d_{s}$ and $y$.  In dimensional regularization, the dimension of space acts as the regulator and becomes a variable.  We denote this analytically continued dimension $d$ to distinguish it from the (fixed) dimension of space $d_{s}$.  Using the usual change of variable $t = \Delta^{2}/(|{\bf p}|^{2} + \Delta^{2})$ and after some algebra the integral becomes:
\begin{eqnarray}
\label{eq:Integral_partial_1}
I(n,d,y) 
\label{eq:Integral_partial_2}
         & = & \left(\frac{2}{(4\pi)^{\frac{d}{2}}}\frac{1}{\Gamma(\frac{d}{2})}\right) \frac{(\Delta^{2})^{-n+\frac{d}{2}-\frac{y}{2}}}{2} \nonumber \\
         &   &\times \left[\frac{\Gamma(-\frac{d}{2}+\frac{y}{2}+n)\Gamma(\frac{d}{2}-\frac{y}{2})}{\Gamma(n)} \right].
\end{eqnarray}
The term in parenthesis comes from geometrical factors and the rest comes from the momentum integral.  In the  white noise limit $y = 0$, the gamma function coming from geometrical factors cancels one coming from the momentum integral, and we are left with the usual result found in textbooks (e.g. \cite{AlvarezGaume_2012}).  None of the complications discussed below arise in that case.

In the case $y \neq 0$, we use the gamma reflection formula $\Gamma(p)\Gamma(1-p) = \pi/\sin \pi p$ to simplify the result.  Assuming $n \geq 1$ we obtain:
\begin{eqnarray}
\label{eq:Main_integral}
I(n,d,y) & = & \left(\frac{2}{(4\pi)^{\frac{d}{2}}}\frac{1}{\Gamma(\frac{d}{2})}\right)\frac{(\Delta^{2})^{-n+\frac{d}{2}-\frac{y}{2}}}{2}\; \frac{1}{\Gamma(n)} \nonumber \\
         &   & \times \frac{\pi}{\sin\pi(\frac{d}{2}-\frac{y}{2})} \prod_{i=1}^{n-1}\left(-\frac{d}{2}+\frac{y}{2} + n - i\right).
\end{eqnarray}
Expression~(\ref{eq:Main_integral}) has poles at $d - y = 2m$, where $m \in {\mathbb{Z}}$.  To study the implications of this, let $d \rightarrow d_{s} + \epsilon$.  The condition becomes $d_{s} - y = 2m - \epsilon$.  Since the noise exponent is a free external parameter dictated by experiment, there is an infinite number of noise exponent values for which this condition is satisfied when $\epsilon \rightarrow 0$ (i.e. for each value of $m$, there is a corresponding value of $y$).    Turning the argument around, it is also possible to ``hit'' different poles by externally varying the noise exponent.  This is an important difference with ordinary quantum field theory.

Another important difference is that the integer $m$ can be either positive or negative.  The case $m > 0$ corresponds to the usual UV divergences encountered in particle physics and they are the focus of the present paper.  The case $m \leq 0$ requires more care.  From Eq.~(\ref{eq:Integral_starting_point}), we see that the integrand is proportional to $d|{\bf p}|\, |{\bf p}|^{2m-1}$.  For $m = -|m|$, the integral measure is (over) compensated by the noise and the integral develops an IR divergence.  This is true even when $\Delta \neq 0$.  Those IR divergences occur for large values of the noise exponent ($y \geq d_{s}$), indicating strong noise correlations at low wavenumbers.  Thus strong noise correlations at large distances are the origin of the IR divergences.

A similar situation arises in thermal field theory, where the temperature acts as an environmental noise.  For definiteness, let us take a massless $\lambda\phi^{4}$ theory at finite temperature.  The one-loop tadpole self-energy is given by (e.g. \cite{Gale_Kapusta_2006}):
\begin{eqnarray}
\Sigma & = & 12\lambda T \sum_{n}\int\frac{d^{3}p}{(2\pi)^{3}}\; \frac{1}{\omega_{n}^{2} + |{\bf p}|^{2}}, \nonumber \\
       & = & 12\lambda \int\frac{d^{3}p}{(2\pi)^{3}}\; \frac{1}{|{\bf p}|}n_{\rm B}(|{\bf p}|),
\end{eqnarray}
where $T$ is the temperature, $n_{B}$ is the Bose-Einstein distribution and the sum is over Matsubara frequencies~$\omega_{n}$.  For $|{\bf p}|/T \ll 1$, the Bose-Einstein distribution can be approximated as ${n_{B}(|{\bf p}|)\sim T/|{\bf p}|}$.  Thus at low energies compared to the temperature, the thermal medium makes the IR behavior of the integrand stronger.  Note that for low energies the Bose-Einstein distribution is like a power-law noise with a negative exponent, similar to our $m \leq 0$ case.  Thus it is plausible to think of the noise as producing IR divergences.  In this paper we do not consider those IR divergences and leave them for future work.

Note that there is no mixing between IR and UV divergences for the stochastic CARD model.  For $m > 0$, the presence of decay rates ($r_{u}$, $r_{v}$) prevents the appearance of IR divergences.  For $m \leq 0$, the momentum powers in the measure are tamed by the noise and cannot produce any UV divergences.  This is in contrast with the renormalization group analysis of the KPZ equation found in Ref.~\cite{Frey_Tauber_1994}.


\subsection{Momentum regularization \label{sec:Momentum_regularization}}

Let us regularize the integral~(\ref{eq:Integral_starting_point}) using both IR and UV momentum cutoffs:
\begin{eqnarray}
\label{eq:Integral_momentum_regularized}
I(n,d_{s},y) & = & \int\frac{d^{d_{s}}\Omega}{(2\pi)^{d_{s}}}\int_{\mu}^{\Lambda} d|{\bf p}|\; \frac{|{\bf p}|^{d_{s}-1-y}}{(|{\bf p}|^{2} + \Delta^{2})^{n}}.
\end{eqnarray}
The result of the integral is:
\begin{widetext}
\begin{eqnarray}
\label{eq:Integral_momentum_regularized_1}
I(n,d_{s},y) & = & \left(\frac{2}{(4\pi)^{\frac{d_{s}}{2}}\Gamma\left(\frac{d_{s}}{2}\right)}\right)\left(\frac{1}{d_{s}-y-2n}\right)\left[\Lambda^{d_{s}-y-2n}{}_{2}F_{1}\left(n,-\frac{d_{s}}{2}+\frac{y}{2}+n,-\frac{d_{s}}{2}+\frac{y}{2}+n+1,-\frac{\Delta^{2}}{\Lambda^{2}}\right) \right. \nonumber \\
             &   & \hspace{1.98in} \left. -\mu^{d_{s}-y-2n}{}_{2}F_{1}\left(n,-\frac{d_{s}}{2}+\frac{y}{2}+n,-\frac{d_{s}}{2}+\frac{y}{2}+n+1,-\frac{\Delta^{2}}{\mu^{2}}\right) \right], 
\end{eqnarray}
\end{widetext}
where ${}_{2}F_{1}(a,b,c;z)$ is the hypergeometric function. The first (second) term corresponds to the UV (IR) divergent part of the integral.  We rewrite Eq.~(\ref{eq:Integral_momentum_regularized_1}) using the series solution representation of the hypergeometric function (see Appendix~\ref{sec:Hypergeometric_function}):
\begin{widetext}
\begin{eqnarray}
\label{eq:Integral_momentum_regularized_2}
I(n,d_{s},y) & = & \frac{1}{2}\left(\frac{2}{(4\pi)^{\frac{d_{s}}{2}}\Gamma\left(\frac{d_{s}}{2}\right)}\right) \left[-\Lambda^{d_{s}-y-2n} \sum_{k=0}^{\infty}\frac{1}{k!}\frac{\Gamma(n+k)}{\Gamma(n)}\frac{1}{\left(-\frac{d_{s}}{2}+\frac{y}{2}+n+k\right)} \left(-\frac{\Delta^{2}}{\Lambda^{2}}\right)^{k} \right. \nonumber \\
             &   &  \hspace{1.26in} -\mu^{d_{s}-y} (\Delta^{2})^{-n} \sum_{k=0}^{\infty}\frac{1}{k!}\frac{\Gamma(n+k)}{\Gamma(n)} 
\frac{1}{\left(\frac{d_{s}}{2}-\frac{y}{2}+k\right)} \left(-\frac{\Delta^{2}}{\mu^{2}}\right)^{-k}  \nonumber \\
      &   & \hspace{1.26in} \left. + \prod_{k}^{n-1}\left(-\frac{d_{s}}{2}+\frac{y}{2}+n-k\right) \frac{1}{\Gamma(n)} \frac{\pi}{\sin\pi\left(\frac{d_{s}}{2}-\frac{y}{2}\right)}  (\Delta^{2})^{\frac{d_{s}}{2}-\frac{y}{2}-n} \right], \\
      & \equiv & I_{1} + I_{2} + I_{3}. \nonumber
\end{eqnarray}
\end{widetext}
Let us study how divergences arise in Eq.~(\ref{eq:Integral_momentum_regularized_2}).  The value of $d_{s}-y$ determines the type of divergence encountered.  Start with $d_{s}-y \notin {\mathbb{Z}}$.  In this case, $I_{3}$ is finite.  For $d_{s}-y > 0$, $I_{2}$ is zero in the $\mu\rightarrow 0$ limit.  When the condition $k_{\rm max} < \frac{d_{s}}{2}-\frac{y}{2}-n$ is satisfied in $I_{1}$, all terms with $k \leq k_{\rm max}$ have positive powers of $\Lambda$.  All other terms (with negative powers of $\Lambda$) vanish in the $\Lambda\rightarrow \infty$ limit.  Thus we are left with a finite number of fractional power UV divergences.  For $d_{s}-y < 0$, $I_{1}$ is finite in the $\Lambda \rightarrow \infty$ limit.  When the condition $k_{\rm max} < \left|\frac{d_{s}}{2}-\frac{y}{2}\right|$ is satisfied in $I_{2}$, all terms with $k \leq k_{\rm max}$ have negative powers of $\mu$.  All other terms (with positive powers of $\mu$) vanish in the $\mu\rightarrow 0$ limit.  Thus we are left with a finite number of fractional power IR divergences.

More care is required for the case $d_{s}-y \in {\mathbb{Z}}$, and one needs to use the limiting procedure $d_{s}-y = 2m - \epsilon$ with $m$ being an integer (c.f. Eq.~(\ref{eq:Definition_limiting_procedure})) to obtain sensible results.  In this case, $I_{3}$ contains $1/\epsilon$ poles.  Such poles also appear in $I_{1}$ or $I_{2}$, depending on the value of $d_{s}-y$.  Since the integral~(\ref{eq:Integral_momentum_regularized}) is regularized and thus finite, all $1/\epsilon$ poles should vanish.  To show that it is indeed the case, let us do an example with $m=0$:
\begin{eqnarray}
\lefteqn{ I(n,m=0) \;\sim\; O(\epsilon^{0}) -\mu^{-\epsilon}(\Delta^{2})^{-n}\frac{1}{\left(-\frac{\epsilon}{2}\right)}}  \nonumber \\ 
  &  &  + \prod_{k=1}^{n-1}(n-k+\frac{\epsilon}{2})\frac{1}{\Gamma(n)} \frac{\pi}{\sin\pi\left(-\frac{\epsilon}{2}\right)}  (\Delta^{2})^{-n-\frac{\epsilon}{2}},  \\
         & = & \frac{2}{\epsilon}(\Delta^{2})^{-n}(1 - \epsilon\ln\mu) - \frac{2}{\epsilon}(\Delta^{2})^{-n} + O(\epsilon^{0}).
\end{eqnarray}
Other values of $m$ are done in a similar way.  Thus for each value of $m$, there is a logarithmic divergence in addition to non-fractional power divergences.

A comparison between Eqs.~(\ref{eq:Main_integral}) and~(\ref{eq:Integral_momentum_regularized_2}) indicates the following.  First note that $I_{3}$ is identical to Eq.~(\ref{eq:Main_integral}) obtained with dimensional regularization.  Thus for ${d_{s}-y = 2m \in {\mathbb{Z}}}$, there is a one-to-one correspondence between $1/\epsilon$ poles in dimensional regularization and logarithmic divergences in momentum regularization.  All other fractional (for ${d_{s}-y \notin {\mathbb{Z}}}$) and non-fractional (${d_{s}-y = 2m \in {\mathbb{Z}}}$) power divergences in momentum regularization are set to zero in dimensional regularization.  This is the expected behavior of dimensional regularization; it is thus a perfectly valid and well-defined way of regularizing integrals in the presence of noise, except for potential fine tuning issues.

In the Standard Model of particle physics, regularizing power divergences to zero using dimensional regularization (for computational purposes) does not make the problem of fine tuning go away.  For instance, quadratic divergences appear when computing loop corrections to the Higgs boson mass $m_{H}$.  This implies that $m_{H}$ is very sensitive to the scale $\Lambda$ at which the Standard Model needs to be replaced by a more fundamental theory (e.g. $\Lambda \sim 10^{16}$ GeV in the case of Grand Unified theories).  Since there is a large hierarchy between the electroweak scale and the scale of new physics, it requires a large fine tuning of the bare mass to keep $m_{H}$ small (e.g. \cite{AlvarezGaume_2012}).  This type of fine tuning is in general considered to be very ``unnatural''.  In our case, the fine tuning is nowhere as severe as in the Higgs boson case.  In complex chemical reactions, there is typically no more than 1-3 orders of magnitude difference in time scales between one set of reactions and the next faster set of reactions (see for example the list of reaction rates of the oscillating Belousov-Zhabotinsky reaction~\cite{Gao_Forsterling_1995}).  This small hierarchy does not require any important fine tuning and thus does not lead to any ``unnaturalness''.

\section{Hypergeometric function \label{sec:Hypergeometric_function}}

The hypergeometric function ${}_{2}F_{1}(a,b,c;z)$ is convergent within the radius $|z| < 1$, but it can be analytically continued to other values of $z$.  The (almost) complete definition of ${}_{2}F_{1}(a,b,c;z)$ is \footnote{See the Wolfram website http://functions.wolfram.com/ HypergeometricFunctions/Hypergeometric2F1/02 for details.}:
\begin{eqnarray}
\label{eq:Definition_2F1_zless1}
{}_{2}F_{1}(a,b,c;z) & = & \sum_{k=0}^{\infty}\frac{(a)_{k}(b)_{k}}{k!(c)_{k}}z^{k},
\end{eqnarray}
for $|z| < 1$ and generic values for $a$, $b$, $c$ and:
\begin{eqnarray}
\label{eq:Definition_2F1_zgreater1}
\lefteqn{{}_{2}F_{1}(a,b,c;z) \;=} \nonumber \\
	&  & \frac{\Gamma(b-a)\Gamma(c)}{\Gamma(b)\Gamma(c-a)} (-z)^{-a} \sum_{k=0}^{\infty}\frac{(a)_{k}(a-c+1)_{k}}{k!(a-b+1)_{k}}z^{-k} \nonumber \\
    &  & + \frac{\Gamma(a-b)\Gamma(c)}{\Gamma(a)\Gamma(c-b)} (-z)^{-b} \sum_{k=0}^{\infty}\frac{(b)_{k}(b-c+1)_{k}}{k!(-a+b+1)_{k}}z^{-k},
\end{eqnarray}
for $|z|>1$ and $a-b \notin \mathbb{Z}$.  When $a-b \in \mathbb{Z}$, we use the limiting procedure:
\begin{eqnarray}
\label{eq:Definition_limiting_procedure}
{}_{2}F_{1}(a,b,c;z) & = & \lim_{\epsilon \rightarrow 0} {}_{2}F_{1}(a,b+\epsilon,c;z).
\end{eqnarray}
In the above:
\begin{eqnarray}
(a)_{k} & = & \left\{\begin{array}{ccc} 1 & & (k=0) \\ a(a+1) \ldots (a+k-1) & & (k>0) \end{array}\right. ,
\end{eqnarray}
is the Pochhammer symbol.

\section{Feynman rules \label{sec:Feynman_rules}}

The following Feynman rules are discussed in Ref.~\cite{Hochberg_etal_2003} (see also~\cite{Barabasi_Stanley_1995} for a more general discussion).  They are obtained by iterating the Fourier-transformed stochastic CARD equations and identifying each component with a picture.  Free response functions are given by (see Fig.~\ref{fig:Feynman_rules}):
\begin{eqnarray}
G_{v0}(k) & = & \frac{1}{D_{v}|{\bf k}|^{2} - i\omega + r_{v}}, \\
G_{u0}(k) & = & \frac{1}{D_{u}|{\bf k}|^{2} - i\omega + r_{u}}.
\end{eqnarray}
They correspond to directed lines, with arrows following the sign of the frequency.  Tree-level interactions are given by (see Fig.~\ref{fig:Feynman_rules}):
\begin{figure}
\includegraphics[width=0.45\textwidth]{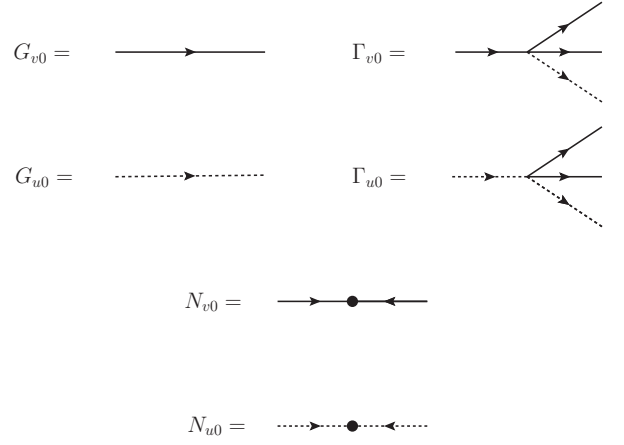}%
\caption{Feynman rules for the stochastic CARD model corresponding to Eqs.~(\ref{eq:Gray_Scott_equations_Fourier_bare_1})-(\ref{eq:Gray_Scott_equations_Fourier_bare_2}).  
\label{fig:Feynman_rules}}
\end{figure}
\begin{eqnarray}
\Gamma_{v0} & = & -\Gamma_{u0} \;\;=\;\; \lambda.
\end{eqnarray}
Without noise, the above response functions and interactions can only produce (arbitrary complicated) tree diagrams (classical theory).  Fluctuations are necessary to obtain loop diagrams in field theory.  The role of fluctuations is played by noise in the stochastic CARD model.  Noise averaging (c.f. Eqs.~(\ref{eq:Noise_property_2})-(\ref{eq:Noise_property_3})):
\begin{eqnarray}
N_{v0}(k) & = & 2A_{v}|{\bf k}|^{-y_{v}}, \\
N_{u0}(k) & = & 2A_{u}|{\bf k}|^{-y_{u}},
\end{eqnarray}
enables the joining of two lines with opposite momenta to form loops (see Fig.~\ref{fig:Feynman_rules}).  Each closed loop corresponds to one noise averaging and thus to one factor of noise amplitude $A$.  In analogy to quantum field theory, the noise amplitude in stochastic partial differential equations plays the role of Planck's constant~\cite{Hochberg_etal_1999}.  A major difference with quantum field theory is that $A$ is an external, tunable parameter whereas $h$ is a constant of Nature.

The components shown in Fig.~\ref{fig:Feynman_rules}, supplemented with conservation of momentum at each vertex and integration over undetermined momenta, form the basis of perturbation theory.  With the appropriate combinatoric factor, they can be used to write down any Feynman diagram for the stochastic CARD model.

\section{One-loop corrections to $\nu$ and $\lambda$ \label{sec:One_loop_corrections}}

From Fig.~\ref{fig:One_loop_corrections} and the Feynman rules in Appendix~\ref{sec:Feynman_rules}, one can write the one-loop correction to the decay rate~$r_{u}$:
\begin{equation}
\Gamma_{r_{u}} = -\lambda\int\frac{d^{d_{s}}p}{(2\pi)^{d_{s}}}\int\frac{d\omega}{(2\pi)}\; G_{v0}(p)G_{v0}(-p)\; 2A_{v}|{\bf p}|^{-y_{v}}.
\end{equation}
Using contour integration to do the frequency integral we get:
\begin{equation}
\label{eq:Correction_ru_after_frequency}
\Gamma_{r_{u}} = -\lambda A_{v} \int\frac{d^{d_{s}}p}{(2\pi)^{d_{s}}}\; |{\bf p}|^{-y_{v}} \left(\frac{1}{D_{v}|{\bf p}|^{2} + r_{v}}\right).
\end{equation}
The above integral is potentially divergent and is regulated using dimensional regularization.  Setting $d_{s} \rightarrow d$ and using formula~(\ref{eq:Main_integral}) to do the momentum integration, we directly obtain Eq.~(\ref{eq:Correction_ru}).

The one-loop correction to the coupling $\lambda$ is done in a similar way.  The expression corresponding to the second diagram in Fig.~\ref{fig:One_loop_corrections} is:
\begin{eqnarray}
\Gamma_{\lambda}(0) & = & -4\lambda^{2}\int \frac{d^{d_{s}}p}{(2\pi)^{d_{s}}}\int\frac{d\omega}{(2\pi)}\; G_{v0}(p)G_{v0}(-p) \nonumber \\
                    &   & \hspace{1in} \times G_{u0}(-p)\; 2A_{v}|{\bf p}|^{-y_{v}},
\end{eqnarray}
where we set external momenta to zero (see the discussion below Eq.~(\ref{eq:Correction_ru})).  Using contour integration to do the frequency integral we get:
\begin{eqnarray}
\label{eq:Correction_after_frequency}
\Gamma_{\lambda}(0) & = & -4\lambda^{2}A_{v}\int \frac{d^{d_{s}}p}{(2\pi)^{d_{s}}}\; |{\bf p}|^{-y_{v}} \left(\frac{1}{D_{v}|{\bf p}|^{2} + r_{v}}\right) \nonumber \\
                    &   & \hspace{0.45in} \times \left(\frac{1}{(D_{u}+D_{v})|{\bf p}|^{2} + r_{u} + r_{v}}\right).
\end{eqnarray}
We combine the two response functions by introducing a Feynman parameter:
\begin{equation}
\Gamma_{\lambda}(0) = \frac{-4\lambda^{2}A_{v}}{D_{v}(D_{u}+D_{v})} \int_{0}^{1} dx  \int\frac{d^{d_{s}}p}{(2\pi)^{d_{s}}} \frac{|{\bf p}|^{-y_{v}}}{\left(|{\bf p}|^{2} + \Delta^{2}(x)\right)^{2}},
\end{equation}
where
\begin{eqnarray}
\Delta^{2}(x) & = & x\left(\frac{r_{v} + r_{u}}{D_{v} + D_{u}}\right) + (1-x)\left(\frac{r_{v}}{D_{v}}\right).
\end{eqnarray}
Equation~(\ref{eq:Correction_lambda}) is obtained by setting $d_{s}\rightarrow d$ and using Eq.~(\ref{eq:Main_integral}).

\begin{acknowledgments}
The authors would like to thank A. Mu\~{n}uzuri,  E. Szabo and J. Szymanski for useful discussions.  J.-S.G. and J.P.-M. thank Repsol S.A. for their support.  DH acknowledges Grant CTQ2013-47401-C2-2-P from MINECO (Spain).
\end{acknowledgments}

\bibliography{UV_properties_GS}

\end{document}